\documentclass[12pt]{article}
\usepackage[utf8]{inputenc}
\usepackage{exscale}
\usepackage{amssymb}
\usepackage{amsmath}
\usepackage{fullpage}
\usepackage{float}
\usepackage{color}
\definecolor{darkgreen}{rgb}{0.0,0.5,0.0}
\definecolor{darkred}{rgb}{0.5,0.0,0.0}
\definecolor{darkblue}{rgb}{0.0,0.0,0.9}
\usepackage{graphicx}
\usepackage{tikz}
\usetikzlibrary{shapes,snakes}
\usepackage{pgfplots}
\pgfplotsset{width=7cm,compat=1.13}
\usetikzlibrary{calc}
\usetikzlibrary[backgrounds]
\usepackage{hyperref}
\hypersetup{
    linktocpage,
     colorlinks,
     citecolor=darkgreen,
     linkcolor= darkgreen,
     urlcolor=darkgreen
}
\usepackage{authblk}
\usepackage{cite}

\def\be{\begin{equation}}
\def\ee{\end{equation}}

\def\tbar{\bar{t}}
\def\bbar{\bar{b}}
\def\pt{p_T}

\def\rsub{R_{\text{sub}}}
\def\fcut{f_{\text{cut}}}
\def\zcut{z_{\text{cut}}}
\newcommand\pythia{\textsc{pythia}}
\newcommand\herwig{\textsc{hewig}}
\newcommand\sherpa{\textsc{sherpa}}
\newcommand\fastjet{\textsc{FastJet}}
\def\MeV{~\text{MeV}}
\def\GeV{~\text{GeV}}
\def\TeV{~\text{TeV}}

\def\msr{\text{MSR}}
\def\mtmc{m_t^{\text{MC}}}
\def\mtp{m_t^{\text{pole}}}
\def\mtfit{m_t^{\text{fit}}}
\def\mwfit{m_W^{\text{fit}}}
\def\mwmc{m_W^{\text{MC}}}
\def\mwpole{m_W^{\text{pole}}}
\def\dmtmc{\Delta m_t^{\text{MC}}}
\def\dmtmctot{\Delta m_t^{\text{MC,tot}}}
\def\dmtfit{\Delta m_t^{\text{fit}}}

\def\mtmsbar{m_t^{\msbar}}
\def\mtmsr{m_t^\msr}

\def\pc{\kappa}

\def\msbar{\overline{\text{MS}}}

\newcommand{\ttbar}{t\bar{t}}

\title{Reducing the Top Quark Mass Uncertainty \\with Jet Grooming}

\author{Anders Andreassen\thanks{anders@physics.harvard.edu} }
\author{Matthew D. Schwartz\thanks{schwartz@physics.harvard.edu}}

\affil{\emph{Department of Physics,
Harvard University, Cambridge, MA 02138, USA}}

\begin{document} 
\maketitle

\begin{abstract}
The measurement of the top quark mass has large systematic uncertainties coming from the Monte Carlo simulations that are used to match theory and experiment. We explore how much that uncertainty can be reduced by using jet grooming procedures. 
We estimate the inherent ambiguity in what is meant by Monte Carlo mass to be around 530 MeV without any corrections. This uncertainty can be reduced by 60\% to 200 MeV by calibrating to the $W$ mass and by 70\% to 140 MeV by additionally
applying soft-drop jet grooming (or to 170 MeV using trimming). 
At $e^+e^-$ colliders, the associated uncertainty is around 110 MeV, reducing to 50 MeV after calibrating to the $W$ mass.
By analyzing the tuning parameters, we conclude that the importance of jet grooming after calibrating to the $W$-mass is to reduce sensitivity to the underlying event. 
\end{abstract}

\newpage


\newpage

\section{Introduction}
The top quark mass is a fundamental parameter in the Standard Model (SM). 
Its value, and the associated uncertainty, are
of great importance for predictions at the Large Hadron Collider (LHC). In the top quark discovery papers from 1995,
the  CDF \cite{Abe:1995hr}  collaboration measured $m_t = 176 \pm 12.8 \GeV$ and D\O~\cite{Abachi:1995iq}  measured 
$m_t =199 \pm 29.7 \GeV$. 
Since then measurements have come a long way, with a recent CMS combination~\cite{Khachatryan:2015hba} using 7 and 8 TeV data giving $m_t = 172.44~\pm~0.48 \GeV$ and a recent ATLAS combination~\cite{Aaboud:2016igd} giving $m_t = 172.84~\pm~0.70 \GeV$. 
Further reducing the uncertainty on the top quark mass is important both for checking self-consistency of the SM and for new physics searches.
For example, because of its order-one coupling to the Higgs, the top quark is a dominant contributor to the Higgs effective potential, with
implications for baryogenesis and vacuum stability.
Indeed, the top quark mass uncertainty is currently the limiting factor in determining whether the Standard Model (SM) is stable or meta-stable \cite{Degrassi:2012ry,Andreassen:2014gha,Espinosa:2015qea}. If $m_t \lesssim 171.22 \GeV$, our universe is unstable, if $m_t \gtrsim 177 \GeV$ it is rapidly unstable. For intermediate values, the universe should last at least as long is it currently has. If $m_t$ is measured precisely enough to confidently claim the Standard Model is in the unstable region this would be compelling evidence for physics beyond the Standard Model. 

For a precision measurement of the top quark mass, we need a precision definition of the top quark mass. 
Since the quark carries color charge, we cannot observe an isolated top quark and measure its mass directly. Instead we have to construct observables that  depend on a top-quark mass parameter in a particular scheme, such as the pole mass, $\msbar$ mass, $1S$ mass~\cite{Hoang:1998hm,Hoang:1999zc},
 potential-subtracted mass~\cite{Beneke:1998rk}, Monte-Carlo mass, $\msr$ mass~\cite{Hoang:2008xm}, etc. (for reviews, see~\cite{Moch:2014lka,Hoang:2014oea,Moch:2014tta}). Then we can fit  the experimental data to a theoretical calculation.
 Some of these schemes, like the $\msbar$ mass, are short-distance mass schemes,
 meaning they are free of renormalon ambiguities and are more stable to the order in perturbation theory at which they are used. 
For the precision in the top-quark mass measurements to continue to improve, understanding the interplay between scheme choice
and experimental uncertainty will be crucial. 
 
The most theoretically sensible way to measure the top quark mass is through an inclusive quantity, like the total $\ttbar$ cross section~\cite{Chatrchyan:2013haa}
or the $\ttbar$ cross section differential in the top $p_T$~\cite{Aad:2015waa}. Such calculations can be performed in perturbative QCD using an unambiguous short-distance mass scheme like $\msbar$. Unfortunately, extractions using cross sections are 
unlikely to produce a top-quark-mass uncertainty below 1 GeV, even at the high-luminosity LHC~\cite{CMS-PAS-FTR-13-017}.
Another approach under good theoretical control  is to look at the production cross section scanning over the energy of the incoming particles, as in $e^+e^- \to \ttbar$~\cite{GUSKEN1985185,Simon:2016htt,Vos:2017ses}.  This method requires a new collider. Using the Large Hadron Collider, the best top quark mass extractions
will come from measurements involving the top quark's hadronic decay products, and therefore it is imperative to get an accurate assesement
of the uncertainty on these methods.

So far, the most precise measurements of the top quark mass have involved fitting the reconstructed top decay products to a theoretical curve. These curves are usually produced using  Monte Carlo (MC) event generators so that the mass scheme used is a Monte-Carlo mass,
$\mtmc$. This Monte-Carlo mass is by definition the value of a parameter in the simulation. It is often assumed to be the same as the pole
mass. To make a precision top-mass measurement, one cannot just assume that $\mtmc = \mtp$, and indeed these two schemes
cannot be the same since $\mtmc$ depends on which Monte-Carlo is used and which tune, while $\mtp$ has a precise field-theoretic definition
(up to a renormalon ambiguity of around 70 MeV~\cite{Beneke:2016cbu}).
Early estimates put the  uncertainty in translating from $\mtmc$ to a well defined short-distance mass scheme like $\msbar$ is of order $1\GeV$ \cite{Hoang:2008xm}, although it seems like the uncertainty may in fact be reducible, perhaps below 100 MeV~\cite{Hoang:2017suc,Butenschoen:2016lpz}.

One approach to  translating the MC mass into a short-distance mass scheme was proposed in~\cite{Butenschoen:2016lpz}. The idea in this paper is to do a precision calculation of an observable related to the top-quark decay products, such as the mass of a highly-boosted top-jet. The calculation should involve a short-distance scheme, and the $\msr$ scheme was preferred~\cite{Hoang:2017suc,Butenschoen:2016lpz}. Then one can fit the distributions from the MC event generators to the theory curves and extract the map from $\mtmsr$ to $\mtmc$. Ideally, one could do these fit in a relatively clean environment, like $e^+e^- \to \ttbar$ events, and the extracted relation between $\mtmc$ and $\mtmsr$ could be applied to values of $\mtmc$ extracted from fits to data at hadron colliders. That is, the program involves two maps: data $\to \mtmc$ and $\mtmc \to \mtmsr \to \mtmsbar$.  The second map seems to be under systematically improvable theoretical control, assuming that the first
map exists, that is, that $\mtmc$ is well-defined. 

In order to use $\mtmc$ for precision mass measurements, one must understand the inherent ambiguity in the definition of $\mtmc$. This ambiguity, related to tuning and limitations of the Monte Carlo programs, contributes to the uncertainty on the extracted top mass and may be the limiting factor in top mass measurements. 
In this paper, we explore how the uncertainty on $\mtmc$ can be reduced, particularly with the use of the jet grooming techniques trimming~\cite{Krohn:2009th} and soft-drop~\cite{Larkoski:2014wba}.

The uncertainty we are concerned with is that the
extracted value of $\mtmc$  can depend on the various parameters of the simulation.
The MC generator has to simulate not only the top quark production and decay, but
initial- and final-state radiation (ISR/FSR), hadronization, secondary interactions in the colliding protons known as either
underlying event (UE) or multiparton interactions (MPI). There is the additional problem of contamination from collisions of other nearby hadrons known as pileup. Pileup  is a stochastic process, uncorrelated with tunings related to $\mtmc$, so we do not consider it here.
By varying the various MC tuning parameters associated with these effects, the same curve (either experimental or theoretical) would match to different values of $\mtmc$, thus we can estimate the uncertainty on $\mtmc$ by varying the tunes. Most experimental top mass measurements provide some estimate of this uncertainty.
For example, the 7 TeV ATLAS top quark mass measurement in the lepton-plus-jets channel~\cite{Aaboud:2016igd} has $\Delta \mtmc = 530 \MeV$, a substantial part of their 1030 MeV total systematic uncertainty from this run. In~\cite{Khachatryan:2015hba} combining 7 TeV and 8 TeV data, CMS estimates
an analogous uncertainty of around 300 MeV.
 There has also been some theoretical work on understanding how different MC parameters, such as the color-reconnection model, effect $\mtmc$~\cite{Skands:2007zg}.

In recent years, a number of jet grooming algorithms have been developed to help clean up jets or events in some way.
Some example groomers  are mass-drop filtering \cite{Butterworth:2008iy}, trimming \cite{Krohn:2009th}, pruning \cite{Ellis:2009me}, modified mass drop~\cite{Dasgupta:2013ihk} and soft drop \cite{Larkoski:2014wba}.
A typical application
is to help resolve subjects in a highly-boosted decay object, like a boosted top quark~\cite{Kaplan:2008ie,Almeida:2008tp} or boosted $W$ boson~\cite{Cui:2010km,Thaler:2010tr}. Another application is to remove radiation from underlying event or pileup so that peaks or shapes in invariant mass distributions are sharper~\cite{Bertolini:2014bba,Krohn:2013lba}. These techniques have shown to be successful in improving signal over background significance. Such applications do not require precision theory: one can find a bump in groomed data without
theory input. Recently, there has been progress in understanding what the groomers are doing 
from perturbative QCD\cite{Dasgupta:2013ihk,Dasgupta:2013via,Dasgupta:2015yua}, 
and it seems promising that precision groomed jet observables might be compared directly to theory without using MC simulations at all~\cite{Frye:2016aiz,Frye:2016okc}.

In this paper, we explore the interplay of jet grooming at the uncertainty on $\mtmc$. 
In Section~\ref{sec:procedure} we describe the setup of our analysis and describe the method used for estimating the uncertainty on the top mass. 
In Section~\ref{sec:jesjetarea} we show
that forcing the reconstructed $m_W$ mass to be exactly $m_W$ is extremely helpful in stabilizing $\mtmc$ over different tunes. 
 Then, in Section~\ref{sec:groom}, we study how grooming techniques trimming and soft-drop can further help reduce uncertainty. In Section~\ref{sec:param}, we explore the parameter space of the groomers and try to get a feel for which parameters are most sensitive to grooming. Our conclusions and a brief discussion is presented in Section~\ref{sec:conc}.

\section{Monte Carlo Top Mass Extraction \label{sec:procedure}}
The basic idea for how we extract the uncertainty on $\mtmc$ is to generate events for each tune for different values of $\mtmc$. Then we fit those distributions to extract a fit mass $\mtfit$. The fit mass will not be the same as the MC mass, but the two are linearly related to an excellent approximation in the regimes we fit: $\mtfit = \pc \, \mtmc$ for some $\pc$. Different tunes give different values of $\pc$ which then translate to an uncertainty on $\mtmc$. More details on the simulation and this extraction procedure are given in this section.

\subsection{Generation of Events}
For our simulated top quark mass measurement we have used the \pythia~8.219 \cite{Sjostrand:2014zea,Sjostrand:2006za} event generator to generate lepton-plus-jet top events, $pp\to t\tbar\to l\nu b\bbar jj$  at $\sqrt{s}=13\TeV$ where $l=e,\mu$. All final state particles (except the neutrinos) with pseudorapidity $|\eta|<4.5$ are clustered using \fastjet~3.2.1 \cite{Cacciari:2011ma} with $\text{anti-k}_{\text{T}}$ \cite{Cacciari:2008gp} with $R=0.5$ (as used by CMS \cite{Khachatryan:2015hba}). 
We require exactly one isolated lepton and at least 4 jets with $\pt>30\GeV$, and that the two $b$-tagged jets are among the 4 jets with highest $\pt$. Only jets with $|\eta|<2.4$ are included in the top reconstruction. 
The lepton and $b$-jets are tagged by matching the four-momentum after the hard interaction to the four-momentum of the jet. If the distance $\Delta R=\sqrt{\Delta \eta^2+\Delta\phi^2}>0.3$ between the four-momentum of the jet and the hard interaction, or if one jet is tagged multiple times, the event is thrown out.  

The events are generated such that $t\to (W^+\to l^+\nu)~b$ and $\tbar \to (W^-\to q'\bar{q})~\bbar$. With the $b$ and $\bbar$ jets tagged, we iterate over all pairs of untagged jets to find the pair with invariant mass closest to $m_W=80.4\GeV$. Only events with the reconstructed $W$ mass between $75\GeV < m_W < 85\GeV$ are kept. The invariant mass of the four-momenta of this pair together with the $\bbar$-jet gives us our reconstructed top quark mass. 

\begin{figure}[t!]
\begin{center}
\begin{tikzpicture}
\node at (0,0) {\includegraphics[width=0.5\columnwidth]{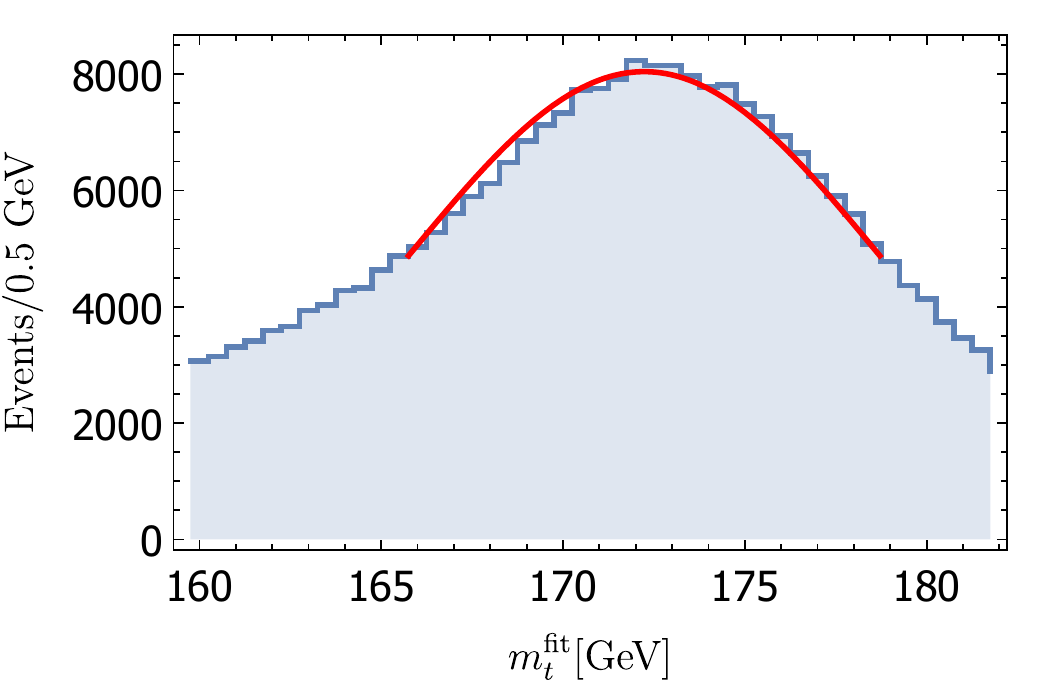}};
\node at (0.9,-0.3) {\includegraphics[width=0.23\columnwidth]{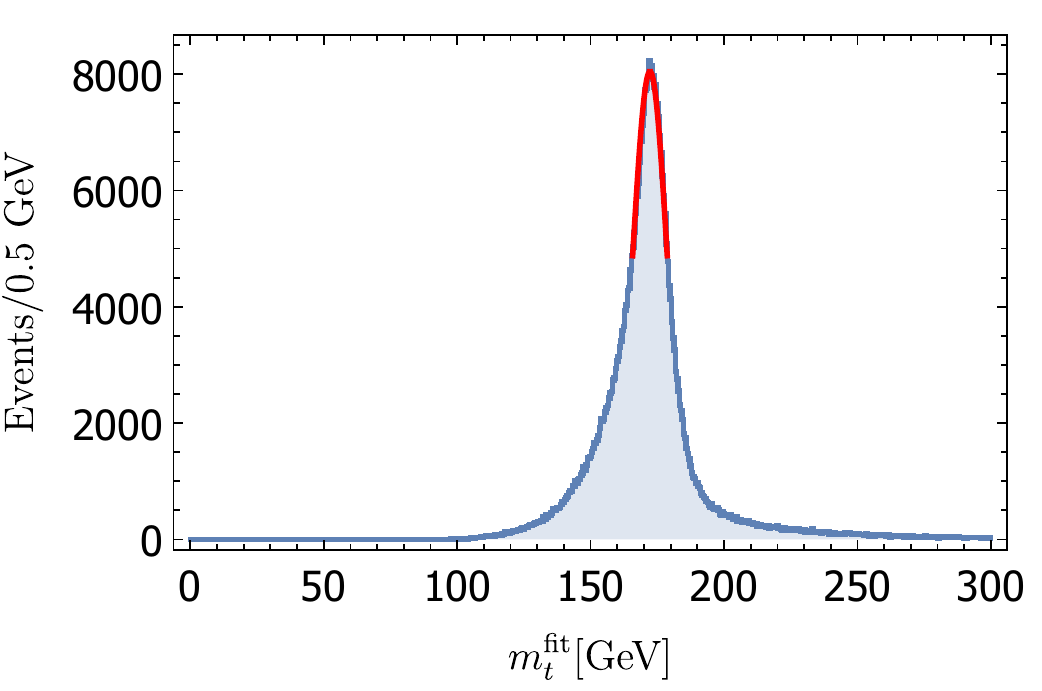}};
\node at (-1.,1.9) {\includegraphics[width=0.22\columnwidth]{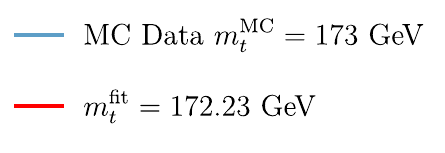}};
\node at (8.,0) {\includegraphics[width=0.5\columnwidth]{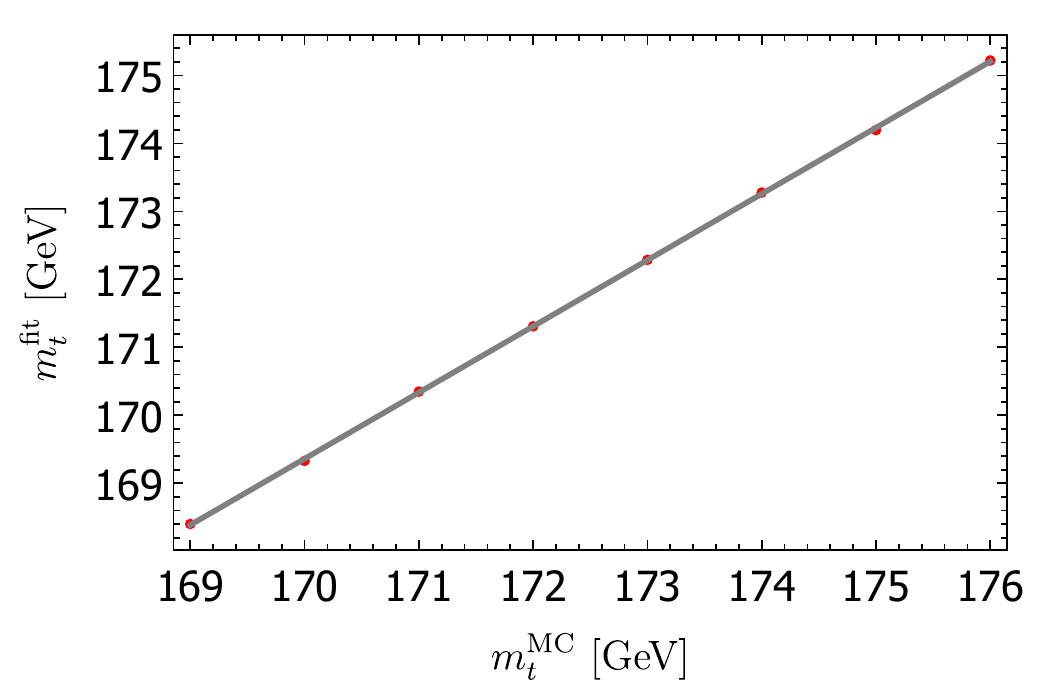}};
\end{tikzpicture}
\caption{Left panel: An example of a histogram of our MC mass measurement, including $W$ calibration (see Section \ref{sec:jesjetarea}), with $\mtmc=173\GeV$.
The plot of the full range from $0-300\GeV$ is inserted. The iterated Gaussian fit gives a simple and good approximation to the peak region. Right panel: The red dots show the calculated value $\mtfit$ for several values of $\mtmc$. A linear fit $\mtfit=0.97\mtmc+3.8\GeV$ shows a good linear relationship.}%
\label{fig:defaultFit}%
\end{center}
\end{figure}

For each run, we generate $10^7$ events of which around 4\% pass our cuts. The reconstructed top quark mass for all events passing the  cuts are then put in a histogram with bin size $0.5\GeV$ that is used for fitting. One such histogram is shown in Fig.~\ref{fig:defaultFit}.  

\subsection{Fitting}
The fitting procedure we used is similar to that implemented by Skands and Wicke in~\cite{Skands:2007zg}.
We fit the simulated data $\frac{d\sigma}{dm}$  to a 3 parameter ($N, \sigma, \mtfit$) Gaussian
\be
f^{N,\mtfit,\sigma}(m)
=N e^{\frac{-(m-\mtfit)^2}{2\sigma}}
\ee
We use a fit range of $ |m - \mtfit| \le \sigma$.
This relatively  narrow window is chosen to avoid sensitivity to the tails of the distribution. 
The fit is done multiple times, each time changing the central value and window, until the fit is symmetric around the peak. 
A typical distribution and fit is shown in Fig.~\ref{fig:defaultFit}. The fit is clearly not perfect, but it does not have to be. One could
get better fits with more parameters, but doing so does not improve extracted uncertainty on $\mtmc$. Indeed, after trying 
more complicated examples, we concluded that a simpler fit gives equivalent results with less statistical variation. 

The goal is to know what the uncertainty on $\mtmc$ is if two tunes give the same value of $\mtfit$.
In the right part of Fig.~\ref{fig:defaultFit}, we show the extraction $\mtfit$ form this procedure for event distributions for a fixed
tuning with different values of $\mtmc$. 
 One can see the relation between $\mtfit$ and $\mtmc$ is linear to an excellent approximation.
$\mtmc = \pc\, \mtfit$. 
If one tune gives $\mtmc= \pc_1 \mtfit$ and another gives  $\mtmc = \pc_2 \mtfit$ then the uncertainty on $\mtmc$ is
$\Delta \mtmc =  \mtfit (\pc_1 - \pc_2)$.

Conveniently, using the linear relation between $\mtmc$ and $\mtfit$, we can skip the the step of varying $\mtmc$ and simply estimate $\pc$ from a single $\mtmc$ value. Suppose for a given $\mtmc$  tune 1 gives $\mtfit{}^{,1}$ and  tune 2 gives $\mtfit{}^{,2}$. Then
we can conclude the second tune would have also  given $\mtfit{}^{,1}$ if
if $\mtmc \to \pc_2 \mtfit{}^{,1}= \mtmc \frac{\mtfit{}^{,1}}{\mtfit{}^{,2}}$. 
Thus the uncertainty on $\mtmc$ is simply
\be
\Delta \mtmc =\mtmc\left(\frac{\mtfit{}^{,1}}{\mtfit{}^{,2}} -1\right) = \Delta \mtfit  \frac{\mtmc }{\mtfit}
\ee
In the following we use this formula we use to estimate $\mtmc$ uncertainty. Note that the factor of $\frac{\mtmc}{\mtfit}$ can be large. For example, for aggressive grooming, we might find $\mtfit \lesssim 0.3\, \mtmc$.

We estimate the uncertainty on this procedure from varying fit shapes and statistical uncertainty in our simulations to be $\Delta \mtfit = 50\MeV$. The effect of assuming a linear relation between $\mtmc$ and $\mtfit$ is negligible.

\subsection{Tunes}
 The parameters in \pythia~are not all independent. In fact, changing parameters separately can result in much more unrealistic events than
changing a handful of parameters in a coordinated way. The recommended way to change simulation parameters
is to vary the \textit{tune}. Each tune in {\pythia} represents  values of the simulation parameters coordinated to give realistic events. 

Choosing which tunes to vary to get a realistic estimate of the MC uncertainty is notoriously subjective. One can choose a subset of tunes and take the envelope of those variations, or one can include the variations from 30 tunes and add the uncertainties in quadrature. The first procedure might underestimate the uncertainty, and the latter probably overestimates it. It is not even clear if all the available tunes span the
possible forms that events could have~\cite{Argyropoulos:2014zoa}. Alternatively, one could vary the simulation itself, comparing {\pythia} to {\herwig} or to {\sherpa} to estimate uncertainties. 

For concreteness, we have chosen to focus our attention on a collection of recent {\pythia} tunes made by ATLAS \cite{ATL-PHYS-PUB-2014-021}, $\texttt{Tune:pp}=19-32$, known as the A14 tunes.  The starting point for these tunes is the Monash 2013 tune \cite{Skands:2014pea}, which is the default tune in \pythia~8.219.
The A14 tunes are divided into six groups (\texttt{Tune:pp} numbers in parenthesis): PDF set variations (19-22), VAR1 (23-24), VAR2 (25-26), VAR3a (27-28), VAR3b (29-30) and VAR3c (31-32).
The PDF set variations correspond to the four different tunes using different PDF sets: CTEQL1 \cite{Pumplin:2002vw}, MSTW2008LO \cite{Watt:2012tq}, NNPDF2.3LO \cite{Carrazza:2013axa} and HERAPDF1.5LO \cite{Sarkar:2014zua}. 
The remaining VAR1-3abc tunes come in pairs with a ``$+$'' and a ``$-$'' systematic variations of the NNPDF tune \cite{ATL-PHYS-PUB-2014-021}. These tunes provides us with a fixed set of tunes that is supposed to cover the range of uncertainties in the MC.
Our overall uncertainty is computed by adding the uncertainty from the six groups of tunes in quadrature.

In addition to using the A14 tunes, we also look at tunes $\texttt{Tune:pp}=14-18$. In some plots, we will show the uncertainty from the envelope over these tunes. We include this as a cross check only; tunes 14-18 are not used to to calculate our overall uncertainty. 
We find the relative reduction in uncertainty using grooming is fairly insensitive to which set of tunes are used, although obviously the absolute size of the reduction does depend on which tunes are chosen. 
We did not look at the comparison with {\herwig} or any other generator, since the procedure for combining the {\herwig} uncertainty with the {\pythia} one is arbitrary. 

For $e^+e^-$, we estimate uncerainty by looking at the envelope over tunes $1,3$ and $7$. 

To be clear, our main concern is the relative improvement in the uncertainty from using grooming. This relative improvement is largely independent of the absolute size of the uncertainty (e.g. soft-drop reduces the uncertainty by 26\%). We quote absolute uncertainties for concreteness, but a proper estimate must be done in the context of the experimental measurement which is beyond the scope of, and not the point of, this paper.

\section{$W$-calibration}\label{sec:jesjetarea}
One of the biggest systematic uncertainties in top mass measurements is due to jet energy scale (JES). For this paper, we define JES as the uncertainty on how much energy and momentum is in a jet given a particular detector response, although other definitions are sometimes used.
One way to calibrate JES is through a standard reference whose
energy is known. For events with top quarks a natural reference is the $W$-boson mass, which is known to precision of a few MeV.
Thus one can demand on an event-by-event basis that the $W$ boson is always reconstructed correctly by rescaling the energy of all particles by some factor~\cite{Skands:2007zg,Argyropoulos:2014zoa}. We call this {\bf ${\mathbf W}$-calibration}. 

$W$-calibration corrects for a lot of issues associated with detector response, so it is common used as a JES correction in experiment. Note however that $W$-calibration also corrects for contamination in the $W$ decay products coming from underlying event, pileup, ISR going into the $W$ decay products and FSR going out of the $W$ decay products. Thus by putting the reconstructed $W$ exactly at the right mass, more than just JES is corrected for. Thus it is meaningful, and indeed very useful as we will see, to use $W$-calibration even for MC-only top mass studies, as we are doing here.

For our implementation of $W$-calibration, we calculate $\mwfit$ from the invariant mass of the $W$ decay products, and 
then we rescale the fit top quark mass by $\mtfit \to \mtfit \frac{\mwmc}{\mwfit}$, with $\mwmc =\mwpole= 80.4\GeV$. 

\begin{figure}[t!]
\begin{center}
\begin{tikzpicture}
\node at (0,0) {\includegraphics[width=1\columnwidth]{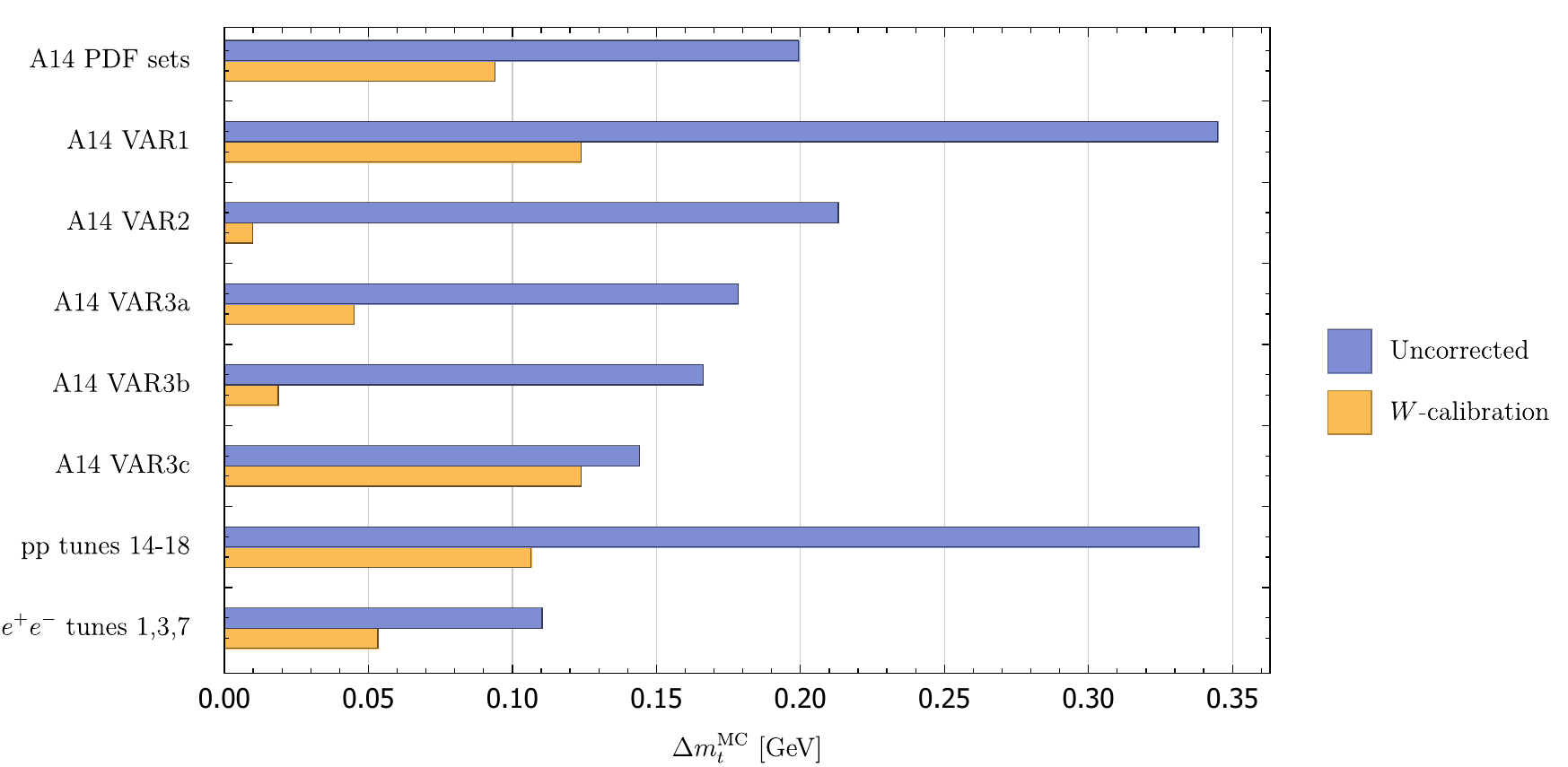}};
\draw[thick,dashed] (-8.3,-1.3) -- (5.1,-1.3);
\draw[thick,dashed] (-8.3,-2.15) -- (5.1,-2.15);
\end{tikzpicture}
\caption{Comparison of $\dmtmc$ for subsets of A14 tunes, ee tunes and pp tunes, with and without $W$ calibration. 
}
\label{fig:A14BarChartJESAREA}
\end{center}
\end{figure}

For each group of A14 tunes, as described in the previous section, we calculate $\dmtmc$ where $\dmtfit$ is taken to be half the difference between the maximum and minimum value of $\mtfit$ within the group. 
The result of $\dmtmc$ with and without the $W$-calibration
 is shown in Fig.~\ref{fig:A14BarChartJESAREA}. 
We find that $W$-calibration significantly reduces the sensitivity to the tune variations, which is what we expected based on the literature of measurements of the top quark mass.\footnote{Skands and Wicke~\cite{Skands:2007zg}  found that $W$-calibration (which they call JES corrections) gave a slight increase of $\dmtmc$ . This is in contradiction to our findings. Details of the variations being studied and improvements in the Monte Carlo simulations over the last ten years make it difficult to reproduce their analysis exactly and may explain the difference.}

Adding the A14 uncertainties in quadrature, $\dmtmctot\equiv \sqrt{\sum_i\left(\Delta m_{t,i}^{\text{MC}}\right)^2}$, we find that with no $W$-calibration $\dmtmctot=530\MeV$, and with $W$-calibration we find $\dmtmctot=200\MeV$. In other words, the combined uncertainty from the A14 tunes is reduced by $62\%$ by including the $W$-calibration. 

In addition to the $W$-calibration, we also tried applying jet area corrections~\cite{Cacciari:2007fd}. This did not lead to any additional improvement.

In Fig.~\ref{fig:A14BarChartJESAREA} we also show the uncertainty coming from the envelope over five other $pp$ tunes. This uncertainty
is smaller than the envelope over the A14 tunes. 

We also show in in Fig.~\ref{fig:A14BarChartJESAREA}  the variations of three $e^+e^-$ tunes.  The uncertainty at $e^+e^-$ colliders  is significantly smaller than the largest uncertainties from the $pp$ tunes (by a factor of 3 without $W$-calibration and a factor of 2 with $W$-calibration). Numerically, the $e^+e^-$ uncertainty is 110 MeV
without $W$-calibration and 50 MeV with $W$-calibration. Keeping in mind that we have estimated around a 50 MeV uncertainty in our fitting procedure, the $W$-calibration has saturated the improvement we can expect for $\mtmc$ at $e^+e^-$ colliders without a more comprehensive study (involving detector simulation, systematic uncertainty and so on, all of which are well beyond the scope of our study). 

\section{Grooming \label{sec:groom}}
In a top mass measurement based on hadronic decay products of the top quark,
 the reconstructed four-momentum of the top is sensitive to the underlying event and initial- and final-state radiation. More underlying event activitiy will typically give a large contribution to the top quark four-momentum, which will directly affect the reconstructed top mass. To mitigate these effects, many different jet grooming algorithms have been introduced to remove wide-angle and/or soft radiation, as mentioned in the introduction. In this section we study how the application of jet grooming techniques can reduce the uncertainty on $\mtmc$. 
We focus our attention on two groomers, trimming~\cite{Krohn:2009th} and soft drop~\cite{Larkoski:2014wba}. Based on the improvements on the systematic uncertainty with $W$-calibration, as seen in the previous section, we will consider both groomed jets with and without the calibration applied. 

\subsection{Optimizing Groomer Parameters\label{sec:optimizing}}
Every grooming algorithm is defined in terms of some set of parameters that we can optimize based on our application. Trimming reclusters each jet using the $k_T$ algorithm \cite{Catani:1993hr,Ellis:1993tq} with characteristic radius $\rsub$, and it discards contributions from subjets which carry less than a  fraction $\fcut$ of the transverse momentum of the original jet. Soft drop reclusters the jet using the Cambridge-Aachen (A/C) algorithm \cite{Dokshitzer:1997in,Wobisch:1998wt}, and depends on two parameters, the soft threshold $\zcut$ and an angular exponent $\beta$. It breaks the jet into two subjets (labeled 1 and 2) by undoing the last stage of the C/A clustering, then checks the soft drop condition $\frac{\text{min}(p_{T1},p_{T2})}{p_{T1}+p_{T2}}=z>\zcut \left(\frac{\Delta R_{12}}{R}\right)^\beta$. If the subjets pass this condition, the jet is the final soft-dropped jet, otherwise the subjet with smaller $p_T$ is thrown out, and the procedure is iterated. 

\begin{figure}[H]
\begin{center}
\begin{tikzpicture}
\definecolor{starcol}{rgb}{1,0,0};
\definecolor{starcolL}{rgb}{1,1,1};
\definecolor{labelcol}{rgb}{0.,0.,0.};
\node at (0,0) {\includegraphics[width=0.42\columnwidth]{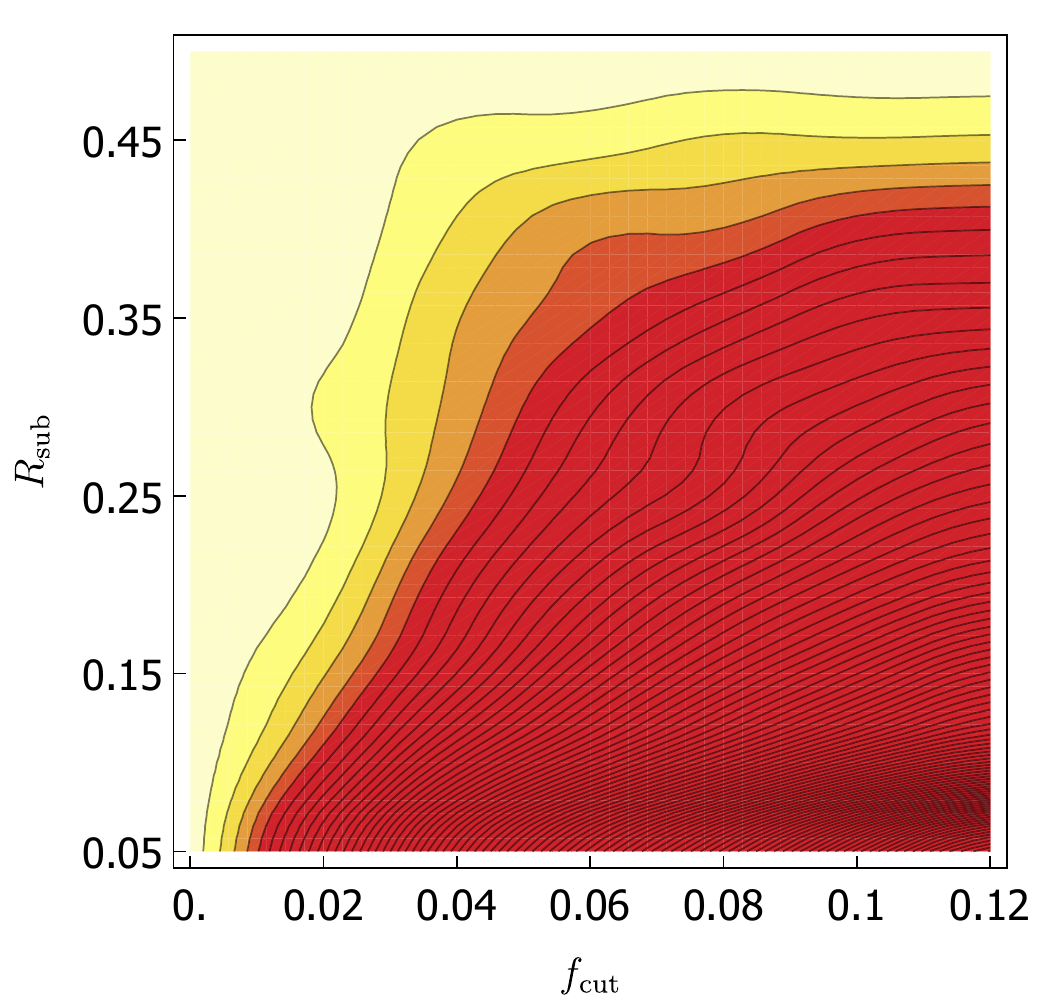}};
\node at (6.7,0) {\includegraphics[width=0.42\columnwidth]{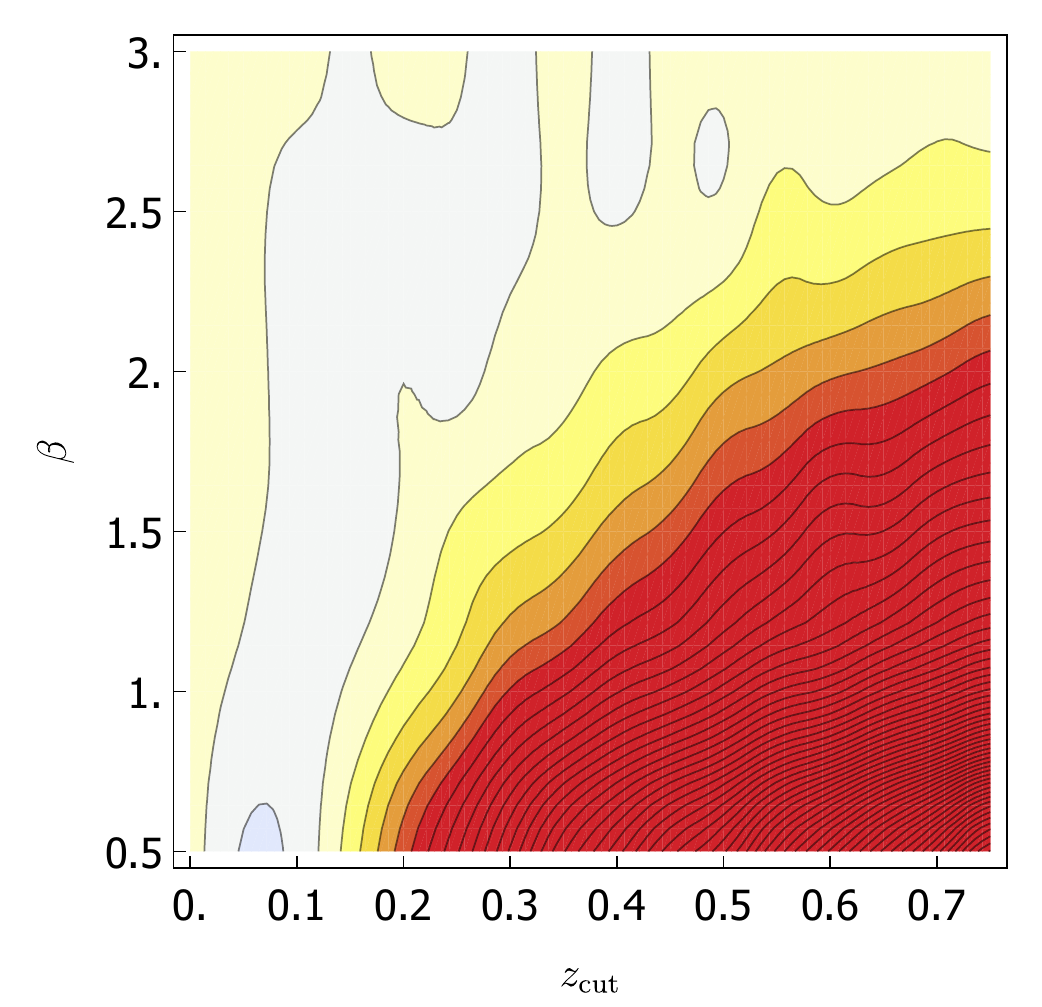}};
\node at (11,0.19) {\includegraphics[width=0.08\columnwidth]{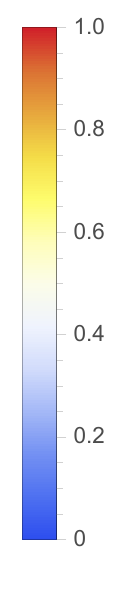}};
\node at (11,3.4) {\tiny{$\dmtmc{}^{\text{,tot}}[\text{GeV}]$}};
\node at (0,-6.6) {\includegraphics[width=0.42\columnwidth]{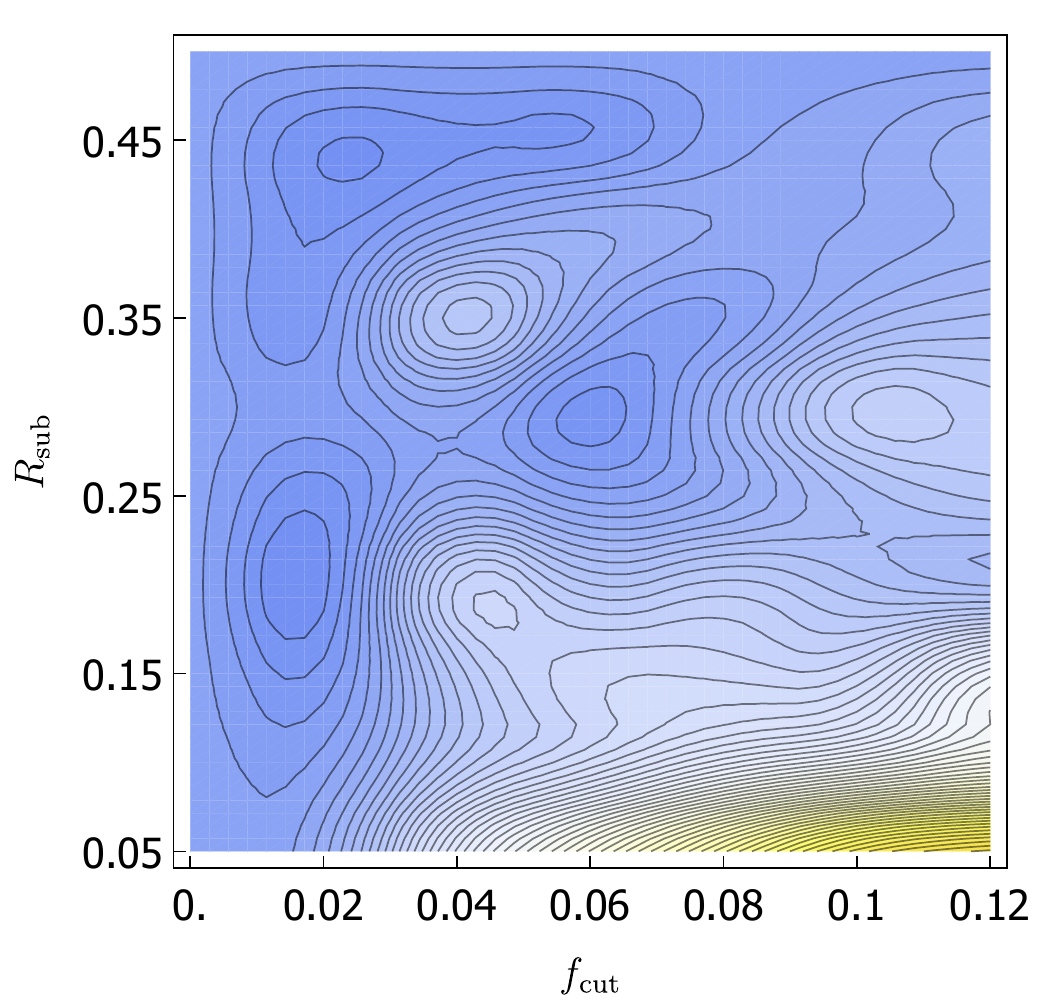}};
\node at (6.7,-6.6) {\includegraphics[width=0.42\columnwidth]{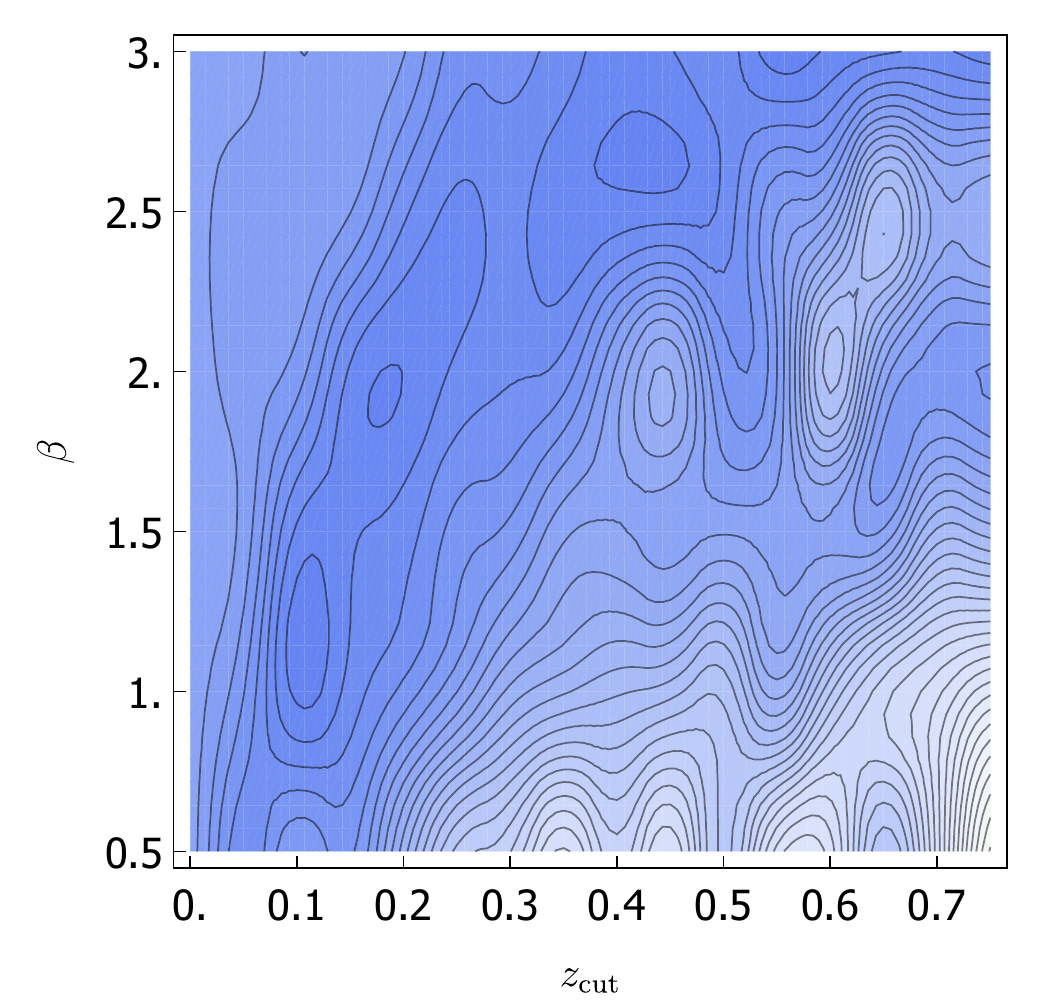}};
\node at (11.,-6.4) {\includegraphics[width=0.08\columnwidth]{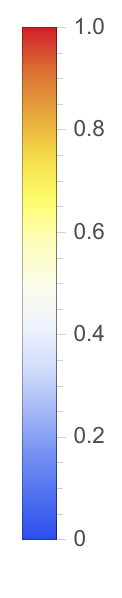}};
\node at (11,-3.2) {\tiny{$\dmtmc{}^{\text{,tot}}[\text{GeV}]$}};
\node[star,star points=5,color=starcolL,fill=starcol,star point ratio=2.25, draw,inner sep=0.01pt] at (5.,-2.3) {$.$};
\node[star,star points=5,color=starcolL,fill=starcol,star point ratio=2.25, draw,inner sep=0.01pt] at (-1.45,-7.1) {$.$};
\node[star,star points=5,color=starcolL,fill=starcol,star point ratio=2.25, draw,inner sep=0.01pt] at (5.28,-7.8) {$.$};
\node[text=labelcol] at (0.5,2.5) {\textbf{Trimming}};
\node[text=labelcol] at (7.25,2.5) {\textbf{Soft drop}};
\node[text=labelcol] at (0.5,-4) {\textbf{Trimming}};
\node[text=labelcol] at (0.5,-4.6) {\textbf{$\mathbf W$-calibrated}};
\node[text=labelcol] at (7.25,-4) {\textbf{Soft drop}};
\node[text=labelcol] at (7.25,-4.6) {\textbf{$\mathbf W$-calibrated}};
\node at (11.2,0.41) { $\leftarrow$};
\node at (12.1,0.45) {\tiny Uncorrected};
\node at (11.2,-6.15) { $\leftarrow$};
\node at (12.1,-6.13) {\tiny Uncorrected};
\end{tikzpicture}
\caption{Contour plot showing $\dmtmctot$ calculated from the six groups of A14 tunes for a range of trimming and soft drop parameters. 
Top panels: without $W$-calibration using contour spacing of 100 MeV; bottom panels: with $W$-calibration using contour spacing of 10 MeV. The stars mark the optimal parameters. There is no star in the first panel since trimming only increases the top-mass uncertainty without $W$-calibration.
}
\label{fig:A14Plots2D}
\end{center}
\end{figure}
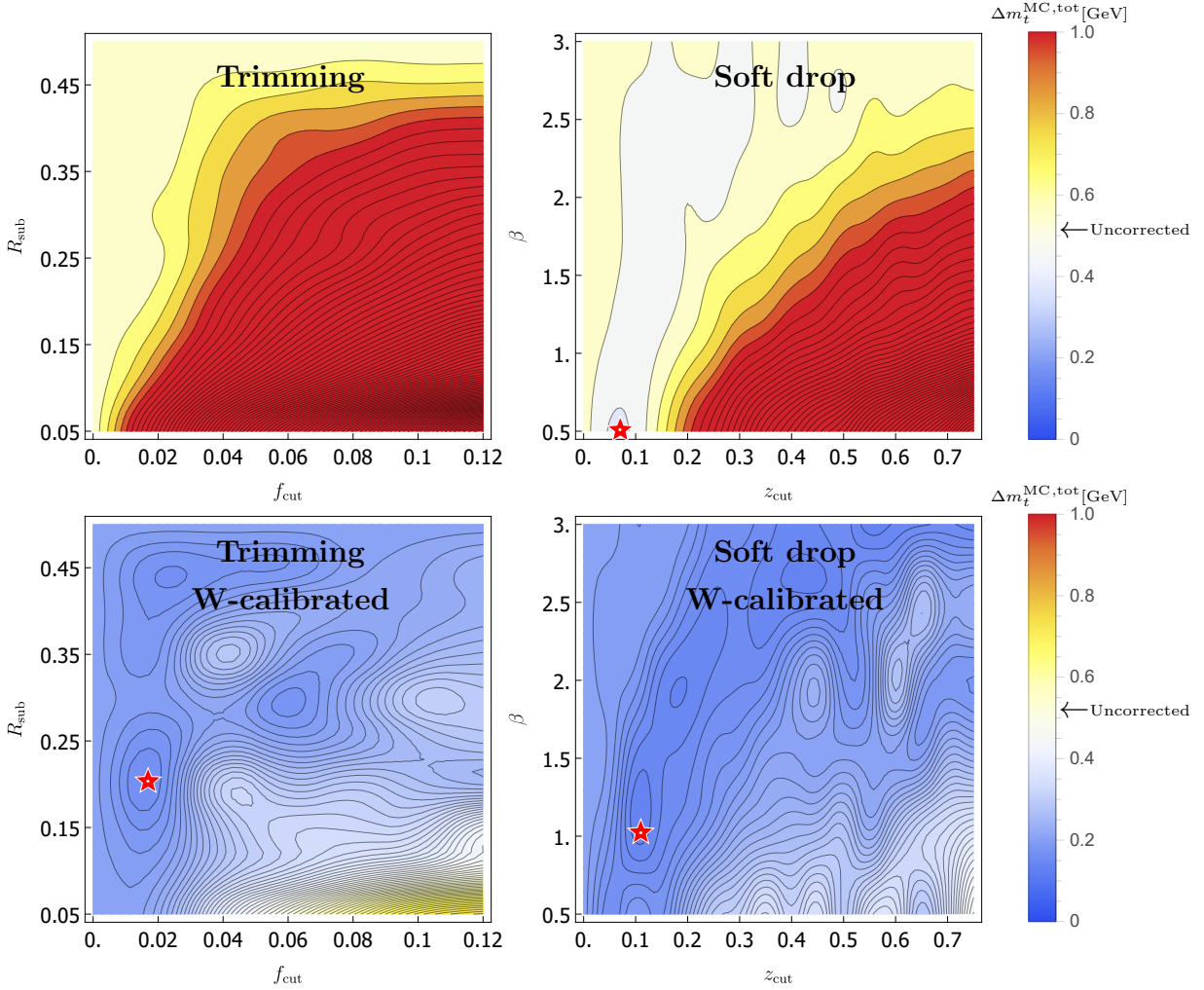

For both trimming and soft drop, we would like to know which grooming parameters minimize $\dmtmc$ as we look at the variations within the A14 tunes. As in Section \ref{sec:jesjetarea}, we will consider the 6 subgroups of the A14 tunes: PDF set variations, VAR1, VAR2, VAR3a, VAR3b and VAR3c. For each group we calculate $\dmtmc$ for each set of groomer parameters, and the uncertainties from the six groups is added in quadrature and 
plotted in Fig.~\ref{fig:A14Plots2D}. 
Without $W$-calibration we find trimming does not help
and for soft-drop  $(\zcut^*,\beta^*)=(0.05,0.5)$ is optimal. 
With $W$-calibration we find for the optimum is at $(\fcut^*,\rsub^*)=(0.02,0.2)$, while for soft drop the optimum is at $(\zcut^*,\beta^*)=(0.1,1.0)$. 
We will call these values our optimized parameters in the rest of this paper. 

\begin{table}[H]
\begin{center}
\caption{Optimal grooming parameters:
\label{tab:optimalgroomingparameters}}
~\\
\begin{footnotesize}
    \begin{tabular}{|l | c | c |}
		\hline
													&\textbf{Trimming}	& \textbf{Soft Drop} \\
													& $(\fcut^*,\rsub^*)$		& $(\zcut^*,\beta^*)$ \\[1mm]
		\hline 
		\textbf{without $W$-calibration} 			& 	--				&  (0.05,0.5)\\[1mm]
		\textbf{with $W$-calibration}& (0.02,0.2)			&  (0.1,1.0)\\
		\hline
	\end{tabular}
\end{footnotesize}
\end{center}
\end{table}

\begin{figure}[ht!]
\begin{center}
\begin{tikzpicture}
\node at (0,0) {\includegraphics[width=1\columnwidth]{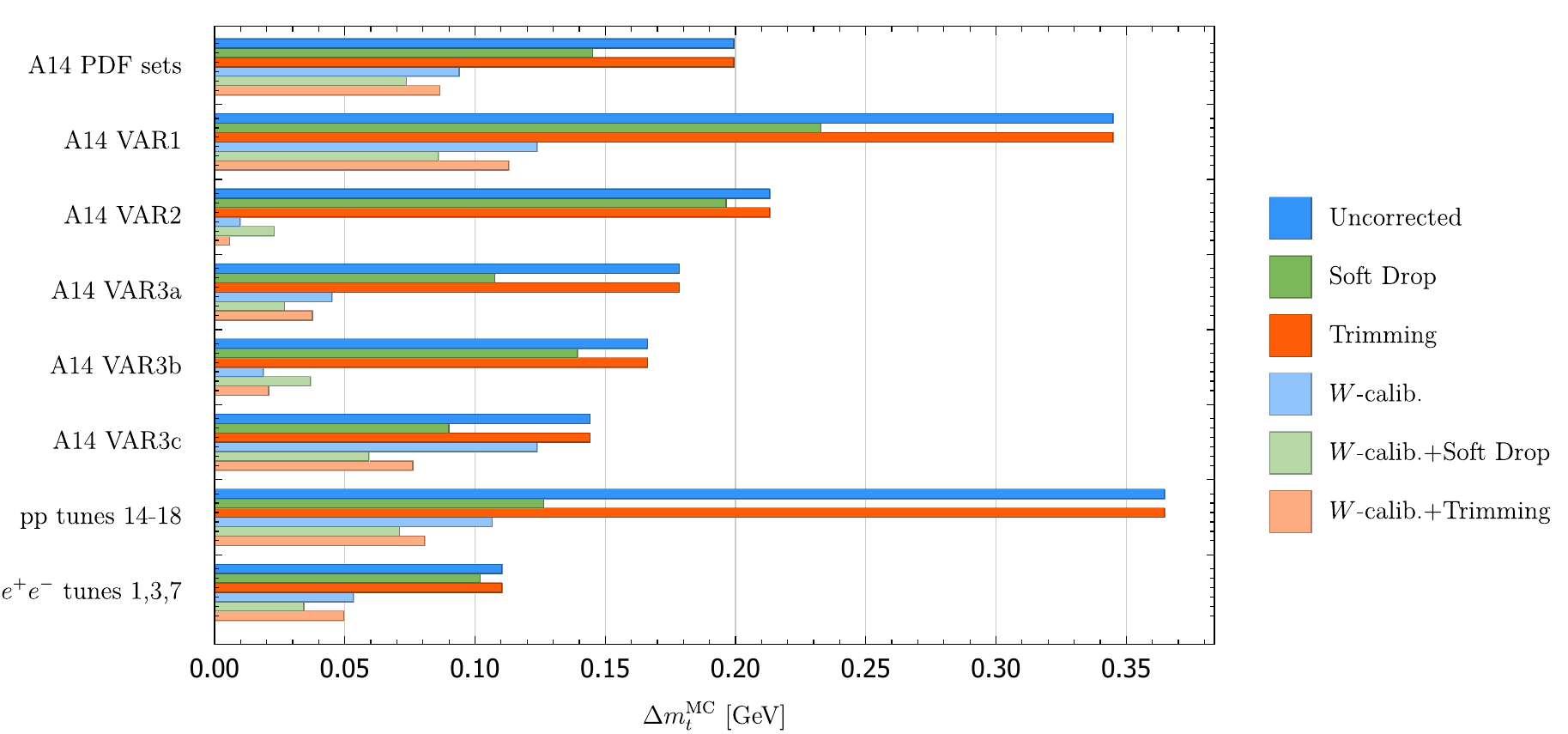}};
\draw[thick,dashed] (-8.3,-1.16) -- (4.2,-1.16);
\draw[thick,dashed] (-8.3,-1.93) -- (4.2,-1.93);
\end{tikzpicture}
\caption{Comparison of $\dmtmc$ for subsets of A14 tunes, pp tunes and $e^+e^-$ tunes, for soft drop, trimming and no grooming for optimized grooming parameters. 
}
\label{fig:A14BarChart}
\end{center}
\end{figure}

\begin{table}[H]
\begin{center}
\caption{Uncertainties on $\mtmc$ after various corrections are included. 
Percentage change from no grooming, without $W$-calibration is shown in parenthesis. 
We estimate around a 50 MeV uncertainty on these numbers due to statistical fluctuations and fitting inaccuracies. 
\label{tab:groom}}
\begin{footnotesize}
~\\
    \begin{tabular}{|l | c c | c c |}
		\hline
														&\textbf{without $W$ calibration} 	&	& \textbf{with W-calibration} &\\[1mm]
													 \hline 
		\textbf{No grooming} 	& $530\MeV$ &				    & $200\MeV$ &$(-62\%)$\\[1mm]
		\textbf{Trimming} 		& $530\MeV$&$(0.0\%)$	& $170\MeV$&$(-68\%)$\\[1mm]
		\textbf{Soft drop} 		& $390\MeV$&$(-26\%)$	  & $140\MeV$&$(-74\%)$\\[1mm]
		\hline
		\textbf{$e^+e^-$} 		  & $110\MeV$&$(-79\%)$	  & $50\MeV$&$(-90\%)$\\
		\hline
	\end{tabular}
\end{footnotesize}
\end{center}
\end{table}

After optimizing the grooming parameters, we study the effect of grooming for each of the A14 groups of tunes. In Fig.~\ref{fig:A14BarChart} we show a comparison of the calculated $\dmtmc$ with soft drop, trimming and no grooming, both with and without $W$-calibration. Our results are summarized in Table~\ref{tab:groom}. In Fig.~\ref{fig:A14BarChart} we also include the uncertainty coming from envelope over tunes
 $\texttt{Tune:pp}=14-18$ (using the A14 optimized groomer parameters). That the uncertainty is in the range of the other tunes indicates that improvements from grooming does not crucially depend on fine tuning of groomer parameters.  We also show the envelope over tunes
 $\texttt{Tune:ee}=1,3,7$ for $e^+e^- \to t\bar{t}$ events. 

For trimming, we see that without $W$-calibration, trimming only makes the uncertainty worse.
After  $W$-calibration, trimming helps in almost all of the tunes.
Adding the A14 tune uncertainties in quadrature, we find that in conjunction with $W$-calibration, the uncertainty is reduced by $68\%$ (compared to 62\% using only $W$-calibration). 

Soft drop helps even without $W$-calibration. With $W$-calibration, it gives an improvement in all variations except VAR2 and VAR3b, whose uncertainties are small anyway.
Adding the uncertainties in quadrature, we find that soft drop gives an improvement of $26\%$ and $74\%$ with and without $W$ calibration, respectively.  These results are summarized in Table~\ref{tab:groom}.

\begin{figure}[t!]
\begin{center}
\begin{tikzpicture}
\node at (0,0) {\includegraphics[width=0.9\columnwidth]{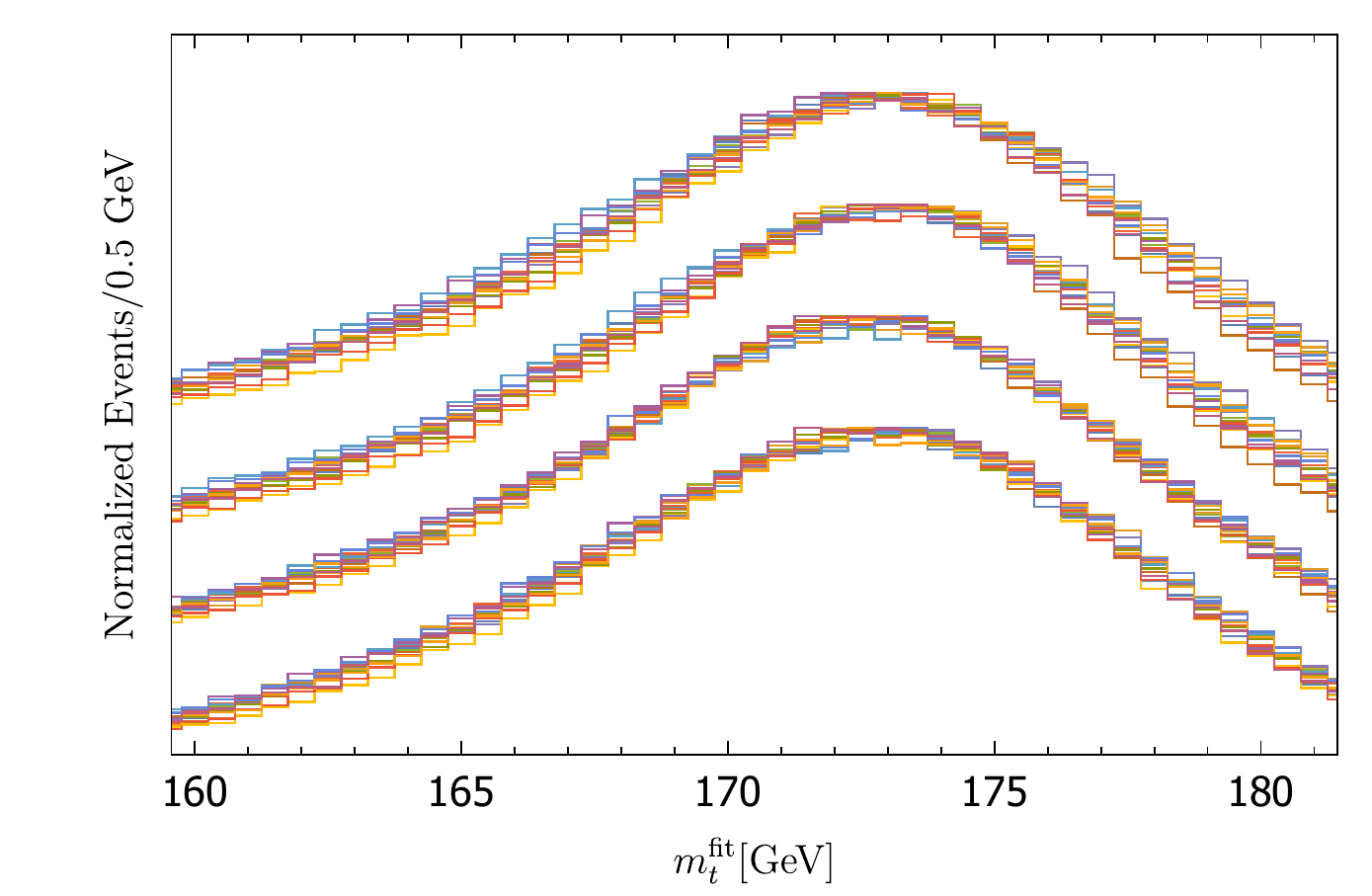}};
\node[] at (2.1,4) {\scriptsize No grooming };
\node[] at (2.1,2.8) {\scriptsize Soft drop only};
\node[] at (2.1,1.55) {\scriptsize $W$-calibration only};
\node[] at (2.1,0.35) {\scriptsize Soft drop \& $W$-calibration};
\end{tikzpicture}
\caption{
Comparison of histograms of $\mtfit$ for the A14 tunes for soft drop and no grooming with and without $W$-calibration for optimized grooming parameters. Each histogram has been normalized (by setting the maximum value to one) and shifted for easier comparison. 
}
\label{fig:CompareHistograms}
\end{center}
\end{figure}
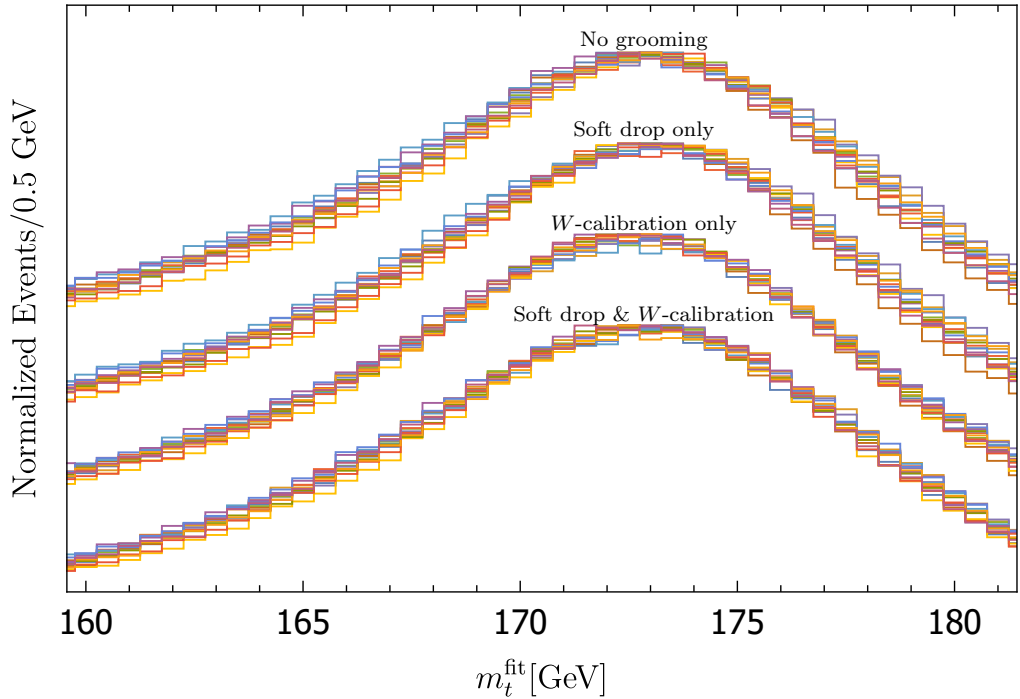

As a cross check, it is informative to look at the shapes of the reconstructed mass distribution in the different cases. These
are shown in Fig~\ref{fig:CompareHistograms}. The $W$-calibration seems to clean up the tails of the distribution. The additional improvement from soft drop seems to improve the peak region slightly, although it is hard to see by eye the origin of the improvement.

\subsection{Changing Individual Parameters \label{sec:param}}
\begin{figure}[ht!]
\begin{center}
\begin{tikzpicture}
\node at (0,0) {\includegraphics[width=1\columnwidth]{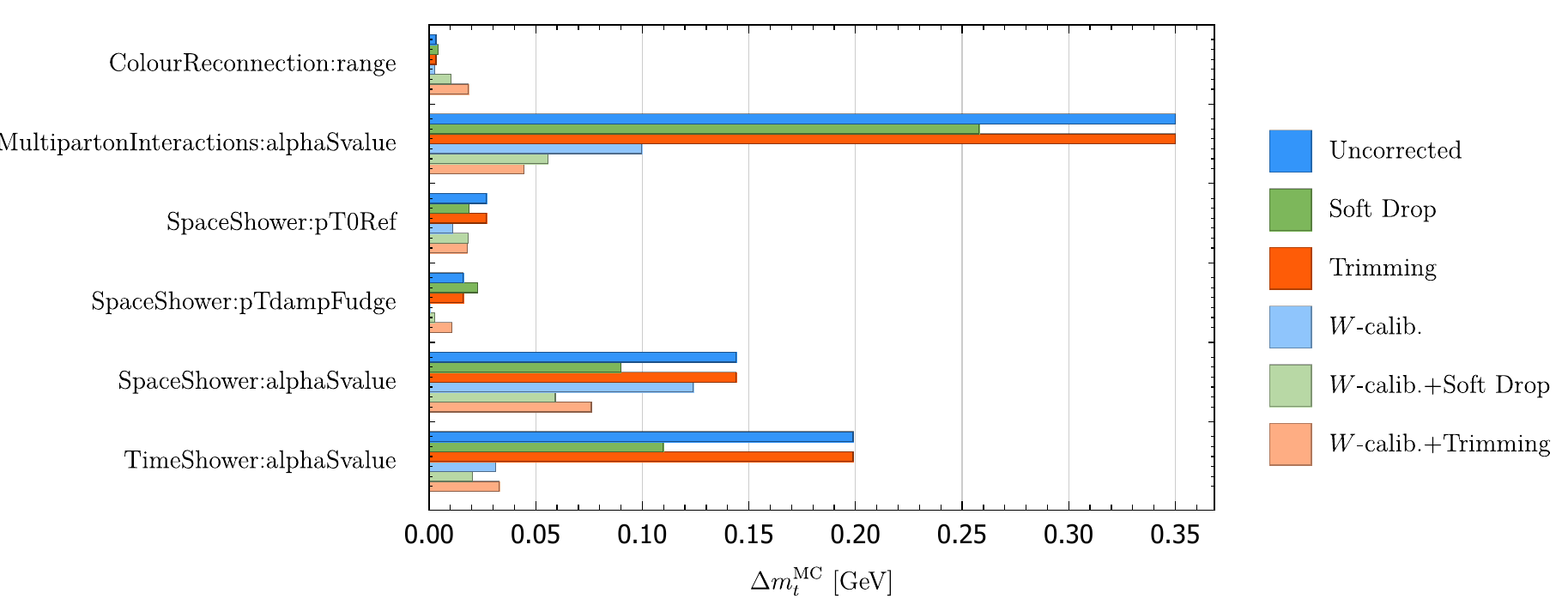}};
\end{tikzpicture}
\caption{Values of $\dmtmc$ comparing the maximum vs. minimum of parameters in Table \ref{tab:A14parameters} starting out with the A14 NNPDF tune for trimming, soft drop and no grooming for optimized grooming parameters $(\fcut^*,\rsub^*)=(0.02,0.2)$ and $(\zcut^*,\beta^*)=(0.1,1.0)$. The \texttt{SpaceShower:pTmaxFudge} results has been omitted as it gave exactly zero variation of the top mass.
}
\label{fig:A14InputSettingsBarChart}
\end{center}
\end{figure}

The A14
tunes contain systematic variations based on the A14 NNPDF tune. 
The parameters that are changed in each tune and the full range of each setting used by these tunes is listed in Table \ref{tab:A14parameters}. In Fig.~\ref{fig:A14InputSettingsBarChart} we show the calculated value for $\dmtmc$ when we compare the maximum vs. minimum value for each individual setting in Table \ref{tab:A14parameters}. Since the variation tunes are based on the A14 NNPDF tune, we set all other parameters to the values described by this tune. Note that changing each setting separately does not give us a good physics description, but it will give us a direct measure of how sensitive the top quark mass measurement is to each of the MC parameters of interest. 

\begin{table}
\begin{center}
\tabcolsep=4pt\relax
\caption{A14 tunes VAR1-3abc parameter ranges}
\label{tab:A14parameters}
\begin{footnotesize}
    \begin{tabular}{| l | c | c | c | c | c | c|}
    \hline
    \textbf{Setting} & \textbf{VAR1} & \textbf{VAR2} & \textbf{VAR3a} & \textbf{VAR3b} & \textbf{VAR3c}   & \textbf{Min-Max}\\ \hline
		\texttt{ColourReconnection:range} & 1.69-1.73 & & & & &1.69-1.73\\
		\texttt{MPI:alphaSvalue} &0.121-0.131  & & 0.127-0.125 & & &0.121-0.131\\
		 \texttt{SpaceShower:pT0Ref} &  & 1.50-1.60& 1.51-1.67& & &1.50-1.67\\
		 \texttt{SpaceShower:pTdampFudge}&  & 1.08-1.04 & 0.93-1.36 &1.07-1.04 & &0.93-1.36 \\
		 \texttt{SpaceShower:pTmaxFudge}&  & & 0.88-0.98 & 0.83-1.00 & &0.83-1.00\\
		 \texttt{SpaceShower:alphaSvalue}&  & & & 0.126-0.129& 0.115-0.140  & 0.115-0.140\\
		\texttt{TimeShower:alphaSvalue} &  & 0.111-0.139&0.124-0.136 &0.138-0.114 & &0.111-0.139\\
		\hline
    \end{tabular}
\end{footnotesize}
\end{center}
\end{table}

Looking at the results in Fig.~\ref{fig:A14InputSettingsBarChart} we find that the dominant uncertainty is coming from the variations of $\alpha_s$ in the multiparton interactions (underlying event), timelike shower (final state radiation) and spacelike shower (initial state radiation).

By comparing the results in Fig.~\ref{fig:A14BarChart} and \ref{fig:A14InputSettingsBarChart} while referencing Table \ref{tab:A14parameters} to see which parameters were varied for each tune, we can understand which settings $\dmtmc$ is most sensitive to.
It is straighforward to see which tuning parameters dominate the uncertainty on the different tunes:
\begin{itemize}
	\item VAR1 is very clearly dominated by \texttt{MultipartonInteractions:alphaSvalue}.
	\item VAR2, VAR3a and VAR3b are dominated by  \texttt{TimeShower:alphaSvalue}, and the size of $\dmtmc$ for each pair of tunes is nicely correlated with the absolute variation of  \texttt{TimeShower:alphaSvalue} for the corresponding pair.
	\item VAR3c is domianted by \texttt{SpaceShower:alphaSvalue}, since this is the only parameter changed.
\end{itemize}

\begin{figure}[h!]
\begin{center}
\begin{tikzpicture}
\definecolor{labelcol}{rgb}{0.,0.,0.0};
\node at (0.8,0) {\includegraphics[width=0.4\columnwidth]{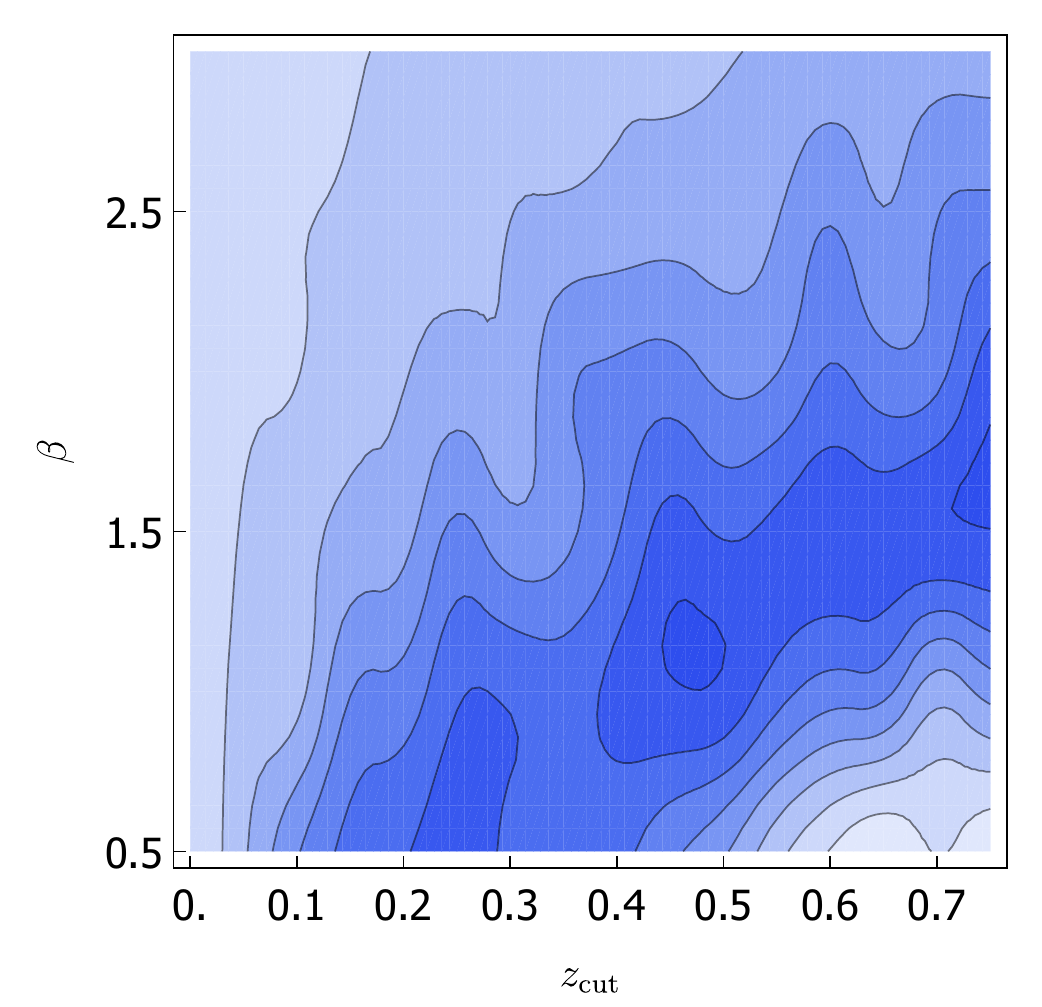}};
\node at (7.45,0.03) {\includegraphics[width=0.4\columnwidth]{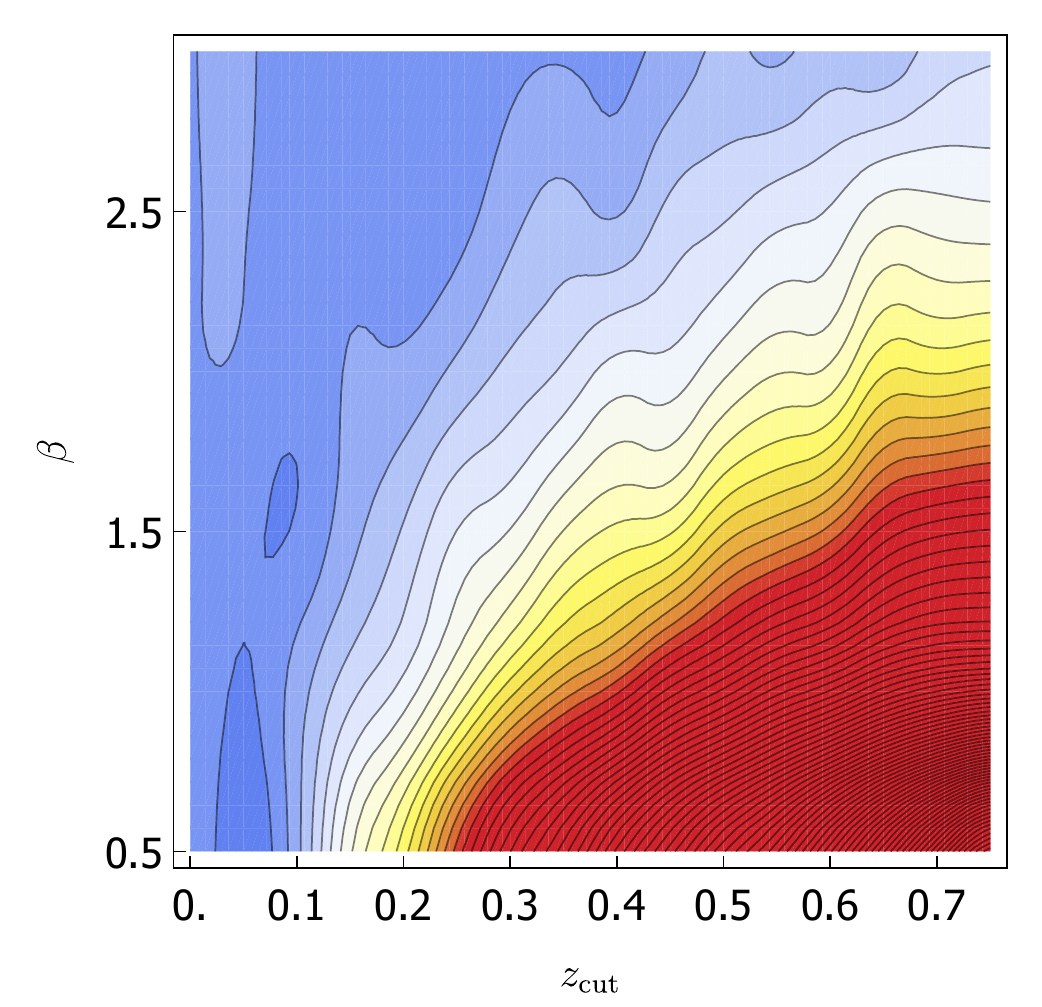}};
\node at (11.1,0.2) {\includegraphics[width=0.075\columnwidth]{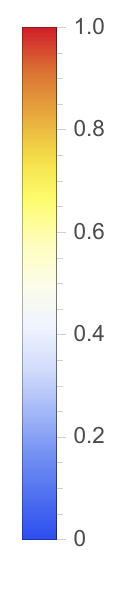}};
\node at (11.1,3.3) {\tiny{$\dmtmc[\text{GeV}]$}};
\node at (1.2,3.2) {\texttt{\footnotesize MultipartonInteractions:alphaSvalue}};
\node at (7.75,3.2) {\texttt{\footnotesize TimeShower:alphaSvalue}};
\node at (0.8,-6.8) {\includegraphics[width=0.4\columnwidth]{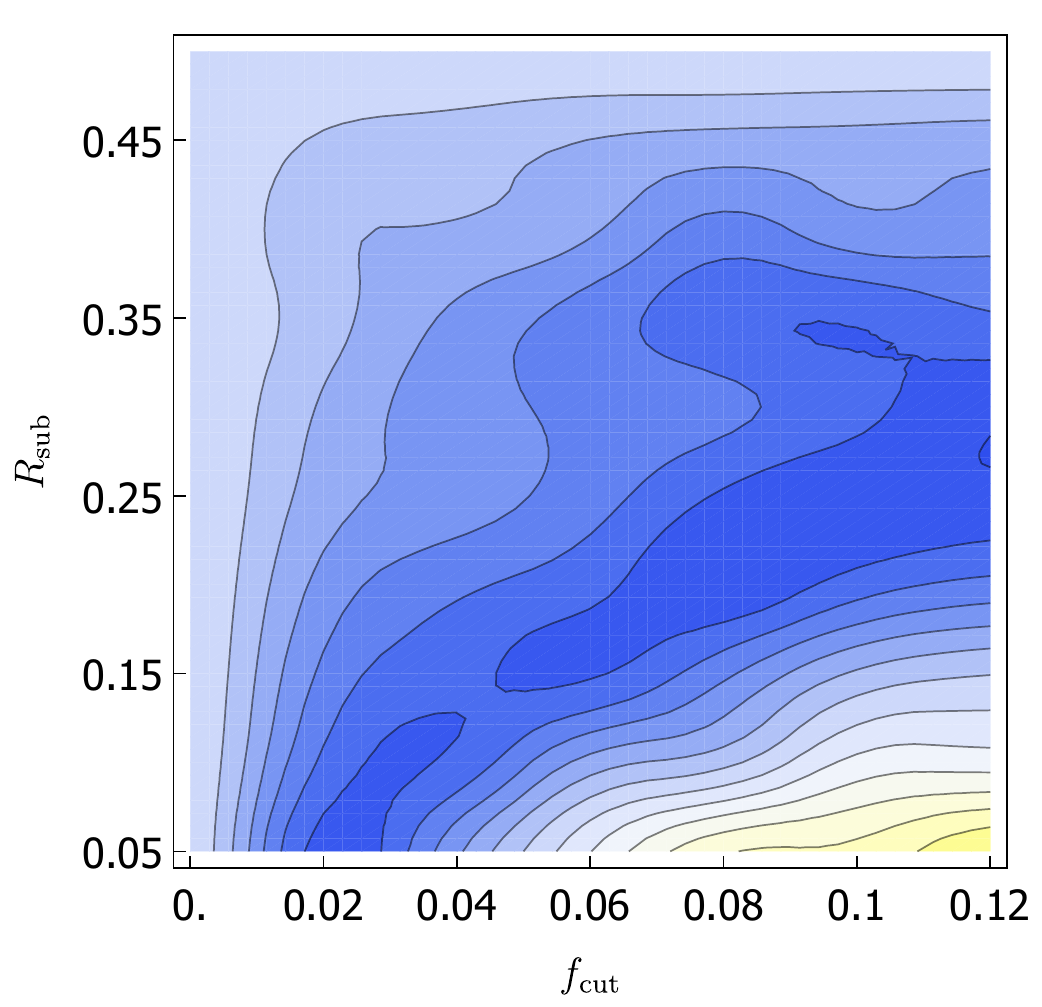}};
\node at (7.45,-6.8) {\includegraphics[width=0.4\columnwidth]{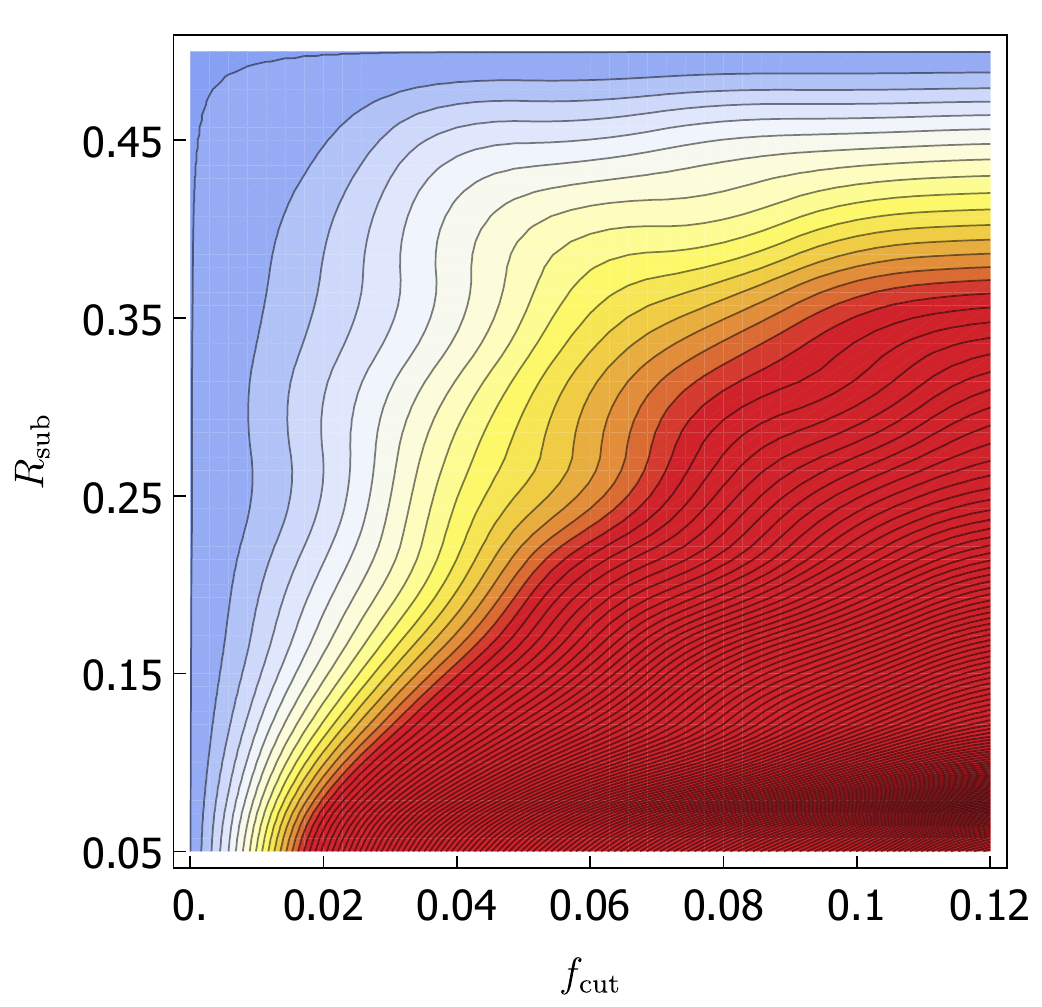}};
\node at (11.1,-6.6) {\includegraphics[width=0.075\columnwidth]{BarLegendA14Settings2DPlot}};
\node at (11.1,-3.6) {\tiny{$\dmtmc[\text{GeV}]$}};
\node at (7.75,-3.6) {\texttt{\footnotesize TimeShower:alphaSvalue}};
\node at (1.2,-3.6) {\texttt{\footnotesize MultipartonInteractions:alphaSvalue}};
\node[text=labelcol] at (1.3, 2.4) {\textbf{Soft drop}};
\node[text=labelcol] at (1.3, 1.8) {\textbf{Underlying event}};
\node[text=labelcol] at (7.9, 2.4) {\textbf{Soft drop}};
\node[text=labelcol] at (7.9, 1.8) {\textbf{Final-state radiation}};
\node[text=labelcol] at (1.3, -4.5) {\textbf{Trimming}};
\node[text=labelcol] at (1.3, -5.1) {\textbf{Underlying event}};
\node[text=labelcol] at (7.9, -4.5) {\textbf{Trimming}};
\node[text=labelcol] at (7.9, -5.1) {\textbf{Final-state radiation}};
\node at (11.3,0.41) { $\leftarrow$};
\node at (12.2,0.45) {\tiny Uncorrected};
\node at (11.3,-6.35) { $\leftarrow$};
\node at (12.2,-6.33) {\tiny Uncorrected};
\end{tikzpicture}
\caption{Contour plot of $\dmtmc$ without $W$-calibration for soft drop and trimming when changing \texttt{MultipartonInteractions:alphaSvalue} and \texttt{TimeShower:alphaSvalue} in the minimum and maximum range listed in Table~\ref{tab:A14parameters}.}
\label{fig:A14Settings2DPlot}
\end{center}
\end{figure}

In Fig.~\ref{fig:A14Settings2DPlot} we show the calculation of $\dmtmc$ (without $W$-calibration) obtained by varying \texttt{MultipartonInteractions:alphaSvalue} and \texttt{TimeShower:alphaSvalue} in the minimum and maximum range listed in Table~\ref{tab:A14parameters} for different grooming parameters. 
For both trimming and soft drop, the grooming is more aggressive as we move towards the lower right corner. Trimming will create many small subjets, and with a higher $\fcut$ it will throw out more and more of them. Looking at the soft drop criterion $z>\zcut\left(\frac{\Delta \theta_{12}}{R}\right)^\beta$, we see that higher $\zcut$ makes it harder to pass the test, and more particles will be thrown out. Also, since $\left(\frac{\Delta \theta_{12}}{R}\right)<1$, smaller $\beta$ will similarly increase the number multiplying $\zcut$, which will make the soft drop condition more difficult to pass.

First consider the case without $W$-calibration. To explain the difference between the MPI and FSR plots, consider the case where we increase $\alpha_s$. In the final state shower, a particle is more likely to split into two (as determined by the splitting functions), and with more aggressive grooming we are more likely to throw out one (or maybe both) of these particles, and hence giving a less accurate reconstruction of the top quark four-momentum. For the multiparton interactions, higher $\alpha_s$ gives more particles produced in the underlying event which in turn contaminate our reconstructed top quark four-momentum. More aggressive grooming (to a certain degree) will hence remove more of the contamination. 

Fig.~\ref{fig:A14Settings2DPlot} therefore confirms our intuition about grooming: aggressive grooming helps remove contamination from the underlying event, but it comes at the expense of throwing out particles that came from the actual decay of the top quark. The optimized grooming parameters strike a balance between the two effects to give the maximal overall improvement of $\dmtmc$.

We can see how things change when including $W$-calibration from Fig.~\ref{fig:A14InputSettingsBarChart}. By putting the $W$ on-shell, problems with aggressive grooming are automatically
compensated for. For example, when we increase $\alpha_s$ for FSR, so that a particle is more likely to split into two smaller subjets and get removed by the groomer, the $W$ mass will also be reduced. Thus the $W$-calibration will compensate for the aggressive groomer in estimating the top mass. Indeed, looking at Fig.~\ref{fig:A14InputSettingsBarChart}, we see that with $W$-calibration included, sensitivity to 
 \texttt{TimeShower:alphaSvalue} in essentially removed, whether or not grooming is additionally applied. 
 \texttt{MultipartonInteractions:alphaSvalue} on the other hand is somewhat reduced, but it is in no way eliminated by the additional $W$ calibration. Thus, Fig.~\ref{fig:A14InputSettingsBarChart} shows that with $W$-calibration, the importance of grooming is to correct for contamination by the underlying event.

\section{Conclusions \label{sec:conc}}
In this paper we have studied the systematic uncertainty of the definition of the top quark Monte-Carlo mass, $\mtmc$. This is the parameter
extracted from experimental fits which so far has given the best top-quark mass measurements.
In order to convert $\mtmc$ to a short-distance mass scheme like $\msbar$, which can be used in precision calculations, one must 
understand the extent to which $\mtmc$ is even well-defined. 
First, there is the question of, for a given MC and a given tune, what short-distance scheme is $\mtmc$ closest to. The traditional answer to this question has been the pole mass $\mtmc \approx \mtp$, since $\mtmc$ appears in hard scattering matrix elements and phase space restrictions just as the pole mass would.  Recently, it has been suggested that in fact $\mtmc$ should be identified with the MSR mass, $\mtmsr$ at a particular scale~\cite{Hoang:2008xm,Hoang:2017suc}. Independent of the conversion to a short-distance scheme, there is the question of the simulation dependence of $\mtmc$. It is the uncertainty on this simulation-dependence that we address in this paper.

Although $\mtmc$ corresponds to a parameter in the Monte Carlo event generator, its extracted value depends on what generator is used and what tune is used within that generator.  By varying the tunes, we found that $\mtmc$ fluctuates by around 530 MeV. A standard experimental procedure to reduce the jet-energy-scale uncertainty is to rescale the energies of the particles so that the $W$-mass is reconstructed exactly. We call this $W$-calibration. 
In addition to mitigating experimental uncertainties associated with detector response,
 $W$-calibration also removes theoretical uncertainties, such as sensitivity to the amount of final-state radiation and underlying event in an event.  We find that by calibrating to the $W$-mass, the uncertainty on $\mtmc$ shrinks to 200 MeV. 

To reduce the uncertainty further, we considered two grooming methods, trimming and soft-drop. We find that on top of $W$-calibration, trimming reduces the uncertainty to 170 MeV while soft drop reduces it to 140 MeV. By looking at the parameters in the different tunes, we saw that the dominant effect corrected by the groomers, but not by the $W$-calibration, is contamination from underlying event. That is, $W$-calibration largely eradicates sensitivity to a dominant source of uncertainty, the amount of final-state-radiation, even before grooming.
In addition, we estimate around a 50 MeV ambiguity on our uncertainties due to the fitting procedure. 

Our estimates were based on adding in quadrature the uncertainties from a set of {\pythia} tunes developed by ATLAS, the A14 tunes. The procedure for calculating theoretical uncertainty is always subjective. Using a different set of tunes, or taking the envelope over the variations rather than adding them in quadrature, or using different MC generators, will all give different absolute numbers. Nevertheless, we believe the relative improvement from $W$-calibration, reducing the uncertainty by about 60\%, and from grooming,  an additional 15-30\% improvement, should be fairly insensitive to the absolute size of the uncertainties. An absolute error estimate is only possible in the context of a particular measurement, including experimental systematic uncertainties, detector effects, and other issues beyond the scope of our study.

We also looked at the analogous uncertainty estimate at $e^+e^-$ colliders. We find without any correction, the uncertainty is around 110 MeV and with $W$-calibration, it reduces to 50 MeV. Since 50 MeV is the same as our estimate of the ambiguity on our fitting procedure, there is no need to consider the effect of grooming on top of $W$-calibration.  

There are two implications of our work. First, we recommend that experimental top mass measurements consider jet grooming in addition to their jet-energy scale corrections. This has the potential to reduce the uncertainty on $\mtmc$ by an additional 30\%. Second, in the pursuit of understanding how to convert $\mtmc$ to a short-distance scheme, like $\msbar$, it will be important to understand the effect of $W$-calibration on theoretical predictions.

\section*{Acknowledgments} 
The authors would like to thank  P. Agrawal, C. Frye, A Hoang, V. Mateu and P. Skands for helpful discussions. This research was supported in part by the Department of Energy under grant DE-SC$0013607$.

\bibliography{TopQuarkMass}

\bibliographystyle{utphys}

\end{document}